\documentclass[10pt,twoside,floatfix,twocolumn,final,superscriptaddress,showkeys,showpacs,aps,prb]{revtex4-1}

\usepackage{graphicx,bm,amsmath,amssymb,color}
\usepackage[utf8]{inputenc}
\usepackage{soul}

\newcommand{\ignore}[1]{}
\newcommand{\mComment}[1]{}
\newcommand{\gComment}[1]{}
\newcommand{\jComment}[1]{}
\newcommand{\rComment}[1]{}
\newcommand{\lComment}[1]{}

\begin{document}
\title{Wave packet dynamics in the one-dimensional extended Hubbard model}

\author{K. A. Al-Hassanieh}
\author{Juli\'an Rinc\'on}
\affiliation{Center for Nanophase Materials Sciences, Oak Ridge National Laboratory, Oak Ridge, TN 37831}

\author{E. Dagotto}
\affiliation{Materials Science and Technology Division, Oak Ridge National Laboratory, Oak Ridge, Tennessee 37831}
\affiliation{Department of Physics and Astronomy, University of Tennessee, Knoxville, Tennessee 37996}

\author{G. Alvarez}
\affiliation{Center for Nanophase Materials Sciences, Oak Ridge National Laboratory, Oak Ridge, TN 37831}
\affiliation{Computer Science \& Mathematics Division, Oak Ridge National Laboratory, Oak Ridge, TN 37831}
\date{\today}

\begin{abstract}
Using time-dependent density-matrix renormalization group, we study the time evolution of electronic wave packets in the one-dimensional extended Hubbard model with on-site and nearest neighbor repulsion, $U$ and $V$, respectively. As expected, the wave packets separate into spin-only and charge-only excitations (spin-charge separation). Charge and spin velocities exhibit non-monotonic dependence on $V$.
For small and intermediate values of $V$, both velocities increase with $V$. However, the charge velocity exhibits a stronger dependence than that of the spin, leading to a more pronounced spin-charge separation. Charge fractionalization, on the other hand, is weakly affected by $V$. The results are explained in terms of Luttinger liquid theory in the weak-coupling limit, and an effective model in the strong-coupling regime.
\end{abstract}

\pacs{71.10.Fd, 71.10.Pm, 75.40.Gb}

\maketitle

\section{Introduction} 

Interacting electrons confined to one-dimensional (1D) geometries exhibit several exotic properties. In higher dimensions, electronic properties can often be successfully described by the Fermi liquid theory of non-interacting quasi-particles. However, this description breaks down in 1D. The low-energy behavior of interacting electrons is universally described by the Luttinger liquid (LL) theory.~\cite{Gogolin,Giamarchi2004} Unlike in the Fermi liquid theory, the low-energy excitations are collective, and obey bosonic statistics.    

One of the most remarkable properties of interacting electrons in 1D is the spin-charge separation (SCS) phenomenon. Luttinger liquid theory predicts that the low-energy excitations decouple into spin (spinons) and charge (holons) excitations that move at different velocities and at different energy scales.~\cite{Gogolin,Giamarchi2004,Deshpande2010} SCS has been observed experimentally in semiconductor quantum wires,~\cite{Auslaender2005} organic conductors,~\cite{Lorenz2002} carbon nanotubes,~\cite{Bockrath1999} and quantum chains on semiconductor surfaces.~\cite{Blumenstein2011} It has also been predicted that SCS can be achieved in optical lattices of ultracold atoms.~\cite{Kollath2005} More recently, SCS has been observed in photoemission experiments on the quasi-1D cuprate SrCuO$_2$,~\cite{Kim} and in electrostatically gated quantum wires deposited on a 2D electron gas.~\cite{Jompol}

In addition to SCS, LL theory predicts charge fractionalization: when an electron (or hole) wave packet with a given momentum is injected into the system, the wave packet splits into left- and right-moving excitations carrying fractional charges. The charge ratio of these fractional excitations is determined by the correlation exponent $K_\rho$ which completely characterizes the LL and 
is defined in terms of the microscopic couplings of the system.~\cite{Giamarchi2004} Charge fractionalization is caused by many-body interactions and the restricted phase space available for scattering in 1D. Indeed, for non-interacting electrons, all the charge moves in one direction, and no fractionalization is expected. Experimentally, charge fractionalization has been observed in quantum wires,~\cite{Steinberg2008} where an imbalance in the currents at the two drains of the wire was detected when an electron with definite momentum was injected in the bulk of the wire.

The basic lattice model used to describe interacting electrons in 1D is the Hubbard Hamiltonian~\cite{Hubbard} which incorporates nearest-neighbor hopping and an on-site Coulomb repulsion. The low-energy properties of this model are well described by the LL theory. Therefore, SCS and charge fractionalization is expected in this model, at least in the long-wavelength limit.~\cite{Gogolin,Giamarchi2004} By resorting to numerical techniques, it has been shown that the SCS persists even beyond the LL regime in finite systems.~\cite{Karen,Ulbricht2009,Jeckelmann,Kollath2005,Julian09,Julian08} Previous studies of the real-time dynamics of wave packets in periodic Hubbard chains~\cite{Karen} showed that the time evolution of charge and spin densities dispersed with different velocities as an immediate consequence of SCS. In strongly correlated models for nanoscopic rings,~\cite{Julian09} numerical calculations of the conductance revealed minima as a function of an applied magnetic flux at fractional values due to SCS. These fractional values correspond to the velocity ratio of spin and charge excitations. The extension of these results to quasi-1D systems has also been considered,~\cite{Julian08} and similar results to the 1D scenario were found. Recent studies on charge fractionalization in the 1D $t$-$J$ model have shown that different fractionalization patterns emerge at higher energy scales.~\cite{Moreno2012}

Studying lattice models beyond the low-energy limit is important because in many materials, electronic interactions fall in the intermediate or strong coupling regimes. In addition, longer range interactions may become relevant leading to new physics with several types of order. Typical materials are ferroelectric perovskites,~\cite{Batlogg88} prototype molecules such as te\-tra\-thia\-ful\-va\-le\-ne-p-chloranil,~\cite{Lemee97} charge transfer organic salts,~\cite{Torrance} organic compounds,~\cite{Rincon09} etc. In this paper, we focus on a simple but interesting extension to the Hubbard model: the addition of nearest-neighbor (NN) repulsion. Such a term gives rise to the so-called extended Hubbard model. This model displays a richer phase diagram~\cite{Voit92,Nakamura00,Jeckelmann2002,Tsuchiizu02,Sandvik04,Batista04,Ejima07,Mund09} and dynamics~\cite{Al-Hassanieh2008, Jeckelmann2000, Bonca2012} than that of the Hubbard model (see discussion below).

Using time-dependent density-matrix renormalization group (tDMRG),~\cite{dmrgreview,tdmrg,Alvarez2011} we study the time-evolution of wave packets created in a 1D extended Hubbard chain with onsite and NN repulsion $U$ and $V$, respectively. 
{\it The results show that $V$ enhances spin-charge separation while weakly affecting charge fractionalization}. 
The main findings can be summarized as follows. (i) Both charge and spin velocities ($v_c$ and $v_s$, respectively) show non-monotonic dependence on $V$. (ii) For small and intermediate values of $V$, $v_c$ increases faster with $V$ and SCS is enhanced. (iii) The results are qualitatively similar in the weak- (small $U$) and strong-coupling (large $U$) regimes, but the underlying mechanisms are different in the two regimes. 

The rest of the paper is organized as follows. Section~\ref{Model} introduces the model and numerical technique. In Sec.~\ref{Strong Coupling} we present a strong-coupling analysis of the model and predict the effect of NN repulsion on SCS. In section~\ref{Weak Coupling} we show the bosonization equations for charge and spin velocities and discuss the effect of $V$ in the weak coupling 
limit. In Sec.~\ref{Results} we present the tDMRG results for the time evolution of spin and charge densities. Final remarks and conclusions are given in Sec.~\ref{Conclusions}.

\section{Model And Technique}
\label {Model}

We investigate an open Hubbard chain of $L$ sites with on-site and NN Coulomb repulsion. The extended Hubbard Hamiltonian is given by

\begin{equation}
\begin{split}
H = &-t_h\sum_{i, \sigma} \left(c_{i\sigma}^\dagger c_{i+1\sigma}^{\;} + \textrm{H.c.}\right) 
+ U\sum_{i} \left(n_{i\uparrow} - \frac{1}{2}\right) \\
&\times\left(n_{i\downarrow} - \frac{1}{2}\right)+ V\sum_{i}\left(n_i - 1\right)\left(n_{i+1} - 1\right),
\label{ham}
\end{split}
\end{equation}
where $t_h$ is the hopping integral and $U$ and $V$ are the on-site and NN repulsion, respectively. The rest of the notation is standard. At half-filling, the ground state (GS) phase diagram of this model is well understood.~\cite{Voit92,Nakamura00,Jeckelmann2002,Tsuchiizu02,Sandvik04,Batista04,Ejima07,Mund09} Both analytical~\cite{Voit92,Nakamura00,Tsuchiizu02,Batista04} and numerical~\cite{Jeckelmann2002,Sandvik04,Ejima07,Mund09} methods have been used to reveal several nontrivial phases of Eq.~(\ref{ham}). The global phase diagram is as follows. For $V<U/2$, the GS of $H$ is a spin density wave (SDW) with antiferromagnetic (AF) quasi-long-range order. For $V>U/2$, the GS is a charge density wave (CDW) with alternating doubly-occupied and empty sites. For small $U$ and $U\sim V$, an intermediate bond-ordered wave state exists for a narrow range of $V$ at the phase boundary between the SDW and CDW phases.~\cite{Batista04} As the SDW state is doped away from half-filling, the holes introduce anti-phase domain walls in the AF order parameter. Similar behavior is expected for the CDW state upon doping. Hence, for slightly doped chains, the SDW-CDW competition is expected to persist.

In our calculations we use $L=50$ sites close to half-filling. The number of electrons per spin is set to $N_\uparrow = N_\downarrow = 23$. The ground state $|\Psi_0\rangle$ is calculated using static DMRG.~\cite{dmrg} Then, a hole Gaussian wave packet with spin $\sigma$ and average crystal momentum $k_0$ is created in the system by applying the operator 
\begin{eqnarray}
h^{\dagger}_{\sigma}(k_0) &=& \sum_j \alpha_j\, e^{ik_0j}\, c_{j\sigma} \nonumber\\
&=& A\sum_j  e^{-(j-j_0)^2/2\sigma_x^2}\, e^{ik_0j}\, c_{j \sigma}
\end{eqnarray}
to $|\Psi_0\rangle$. The coefficients $\alpha_j$ are chosen to generate a Gaussian wave packet centered at site $j_0$ with width $\sigma_x$. The prefactor $A$ is a normalization constant. In momentum space, the wave packet is centered at momentum $k_0$ with width $\sigma_k = 1/\sigma_x$. We use $k_0 = 0.46\pi - 4 \sigma_k$ and $\sigma_k = 0.075$; i.e.~$k_0$ is set close to the Fermi 
momentum which lies within the linear dispersion regime.  

The excited state $|\Psi_e\rangle = h^{\dagger}_{\uparrow}(k_0) |\Psi_0\rangle$ is time-evolved under $H$: $|\Psi(t)\rangle = e^{-iHt} |\Psi_e\rangle$, where $H$ is given by Eq.~(\ref{ham}). We follow the time evolution of the charge and spin densities $n_i(t) = n_{i\uparrow}(t) + n_{i\downarrow}(t)$ and $s_i(t) = (n_{i\uparrow}(t) - n_{i\downarrow}(t))/2$, respectively, with 
\begin{equation}
n_{i\sigma}(t) = \langle \Psi(t)|n_{i\sigma}|\Psi(t)\rangle - \langle \Psi_0 |n_{i\sigma}|\Psi_0\rangle
\end{equation}
 so that in the results shown, the GS expectation values are subtracted from the time-dependent measurements. By performing this subtraction, we avoid characteristic oscillations of the GS of the system. The algorithm implemented for the time-dependent part of DMRG is based on the Krylov space decomposition algorithm. An implementation can be found in Refs.~\onlinecite{Alvarez2009, Alvarez2011}. In the time-evolution  calculations, 600-800 states per block are kept. The results are well converged with this number of states.

\section{Strong Coupling Limit}
\label{Strong Coupling}


We present first an intuitive picture of $H$ in the strong-coupling limit $U\gg t_h$ and predict the effect of $V$ on SCS. For $U\gg t_h$ and $V=0$, the low-energy physics of $H$ is described by an effective $t_h$-$J$ model with AF coupling $J=4t_h^2/U$. The holes 
move with hopping $t_h$. In this limit, the charge and spin velocities are determined by the holon and spinon bandwidths, respectively. $v_c \propto t_h$ and $v_s \propto J$ . For finite but small $V$ ($V< U/2$), a qualitatively similar behavior is expected. In order to understand the effect of $V$ quantitatively, we discuss the charge and spin sectors separately.

In the spin sector, it is easy to see that $V$ renormalizes the AF spin coupling to $J = 4t_h^2/(U-V)$. Thus $J$, and consequently $v_s$, increase slowly as $V$ increases in the range $V<U/2$. This is observed in the numerical results. More precisely, using Bethe-ansatz results for the strong-coupling limit,~\cite{Shiba72} we have for the spin excitations velocity
\begin{equation}
v_s = \pi J\left(1-\frac{\sin{4k_F}}{4k_F} \right) \approx \pi J\left(1+\delta \right),
\label{vsSC}
\end{equation}
where we have considered a slightly doped half-filled band with $2k_F = \pi(1-\delta)$, and $\delta \ll 1$ is the density of holes.

The effect of $V$ is more pronounced in the charge sector. We start with the $V=0$ case in which the model can be mapped into the AF $t_h$-$J$ model as described above. To simplify the discussion and in order to focus on the charge sector, we assume that the spin interaction is Ising-like, i.e., we consider a $t_h$-$J_z$ model. This model can be mapped exactly into a spinless fermions model with hopping $t_h$ and NN attraction $J_z/4 = t_h^2/U$.~\cite{Batista2000} The effect of $V$ is easy to see in this model. When $V>0$, the NN interaction between spinless fermions becomes $V-J_z/4 = V-t_h^2/U$. The bosonization analysis of this model gives the charge velocity~\cite{Gogolin,Giamarchi2004}
\begin{equation}
v_c= v_F\sqrt{1+\frac{2}{\pi}\frac{V-t_h^2/U}{v_F}\left(1-\cos{2k_F}\right)},
\end{equation}
where $v_F=2t_h\sin{k_F}$. For a band close to half-filling in the extended Hubbard model $H$, $2k_F = \pi(1-\delta)$. The charge velocity can then be written as 
\begin{equation}
v_c \approx v_F \sqrt{1+\frac{2}{\pi}\frac{V-t_h^2/U}{v_F}\left(2-\frac{\pi^2\delta^2}{2}\right)}.
\label{vcSC}
\end{equation} 
These equations show that $v_c$ increases with $V$ in the strong-coupling limit faster that the spin velocity $v_s$ shown in Eq.~(\ref{vsSC}).


Based on the above arguments, $v_c$ and $v_s$ have a non-monotonic dependence on $V$ with a maximum around intermediate $V$. SCS is enhanced for small $V \approx 2t_h$.

\begin{figure}
\centering
\includegraphics*[width=\columnwidth]{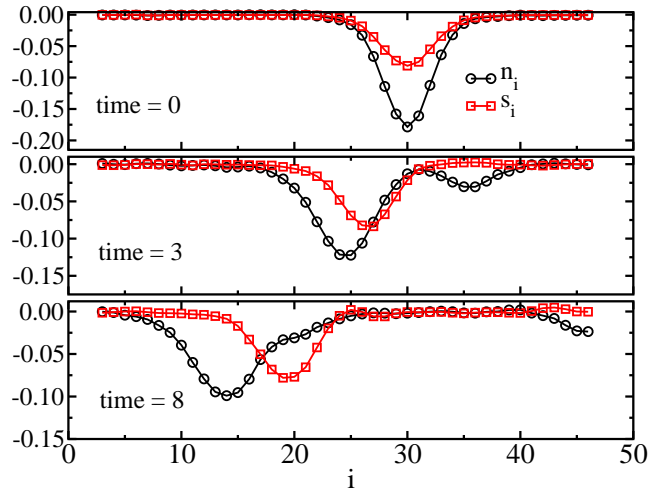}
\caption{Charge and spin propagation in the weak-coupling limit ($U=2t_h$ and $V=0$). $n_i$ (circles) and $s_i$ (squares) are shown at different times. At $t=0$, the wave packet with charge and spin is created centered at $j_0=30$. The spin and most of the charge move in the same direction but at different velocities. SCS is evident starting at $t = 3$, as is the charge fractionalization since a small fraction of the charge moves in the opposite direction. As expected in this limit, charge fractionalization is small (see text).}
\label{Fig1}
\end{figure}

\section{Weak Coupling Limit}
\label{Weak Coupling}

In the weak coupling limit, $U,V\ll v_F$, we can resort to bosonization formulas for charge and spin velocities in the presence of $U$ and $V$. These couplings give rise to different effective interactions in each sector: $U_c$ and $U_s$.~\cite{Giamarchi2004} The expressions for the velocities are 
$v_c = v_F (1+U_c/\pi v_F)^{1/2}$,
and 
$v_s = v_F (1-U_s/\pi v_F)^{1/2}$. 
The weak-coupling interactions are given by $U_c = U + 2V(1-2\cos(2k_F))$ and $U_s = U + 2V\cos(2k_F)$.

\begin{figure}
\includegraphics*[width=\columnwidth]{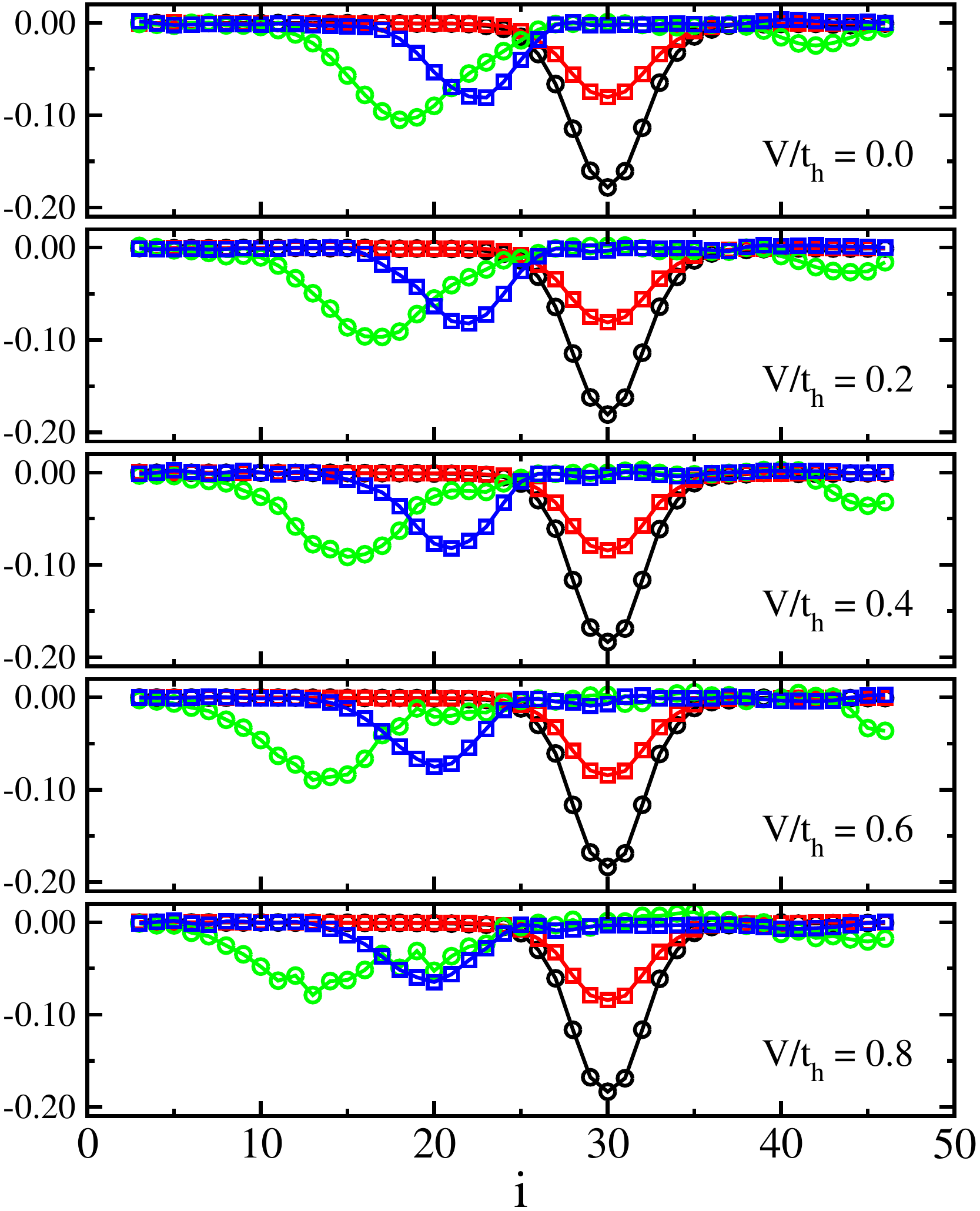}
\caption{Charge and spin propagation in the weak-coupling limit ($U = 2t_h$) for different values of $V$. For each $V$, $n_i$ (circles) and $s_i$ (squares) are shown at times $t=0$ and $t=6$. For $V/t_h \le 0.6$, both $v_c$ and $v_s$ increase with $V$ (see Eqs.~(\ref{vsWC}) and (\ref{vcWC})). $v_c$ has a stronger dependence on $V$ and SCS is enhanced (see Eq.~(\ref{vcWC})). Charge fractionalization, on the other hand, is weakly affected by $V$. As $V$ approaches $U/2$, both velocities decrease and SCS is suppressed.}
\label{Fig2}
\end{figure}

Close to half-filling, the equations simplify to 
\begin{equation}
v_s \approx v_F\sqrt{1-\frac{U}{\pi v_F}+\frac{2V}{\pi v_F}\left(1-\frac{\pi^2\delta^2}{2}\right)}
\label{vsWC}
\end{equation}
and
\begin{equation}
v_c \approx v_F\sqrt{1+\frac{U}{\pi v_F}+\frac{2V}{\pi v_F}\left(3-\pi^2\delta^2\right) }.
\label{vcWC}
\end{equation}
Thus, the velocities for both excitations increase with $V$ and, as in the case of the strong-coupling limit, they are only slightly affected by doping effects. SCS is then enhanced for small $V$. For the large $V$ regime, the ground state exhibits CDW, and SCS is suppressed as in the strong-coupling limit (see below).

\section{Results}
\label{Results}

\begin{figure}
\includegraphics*[width=\columnwidth]{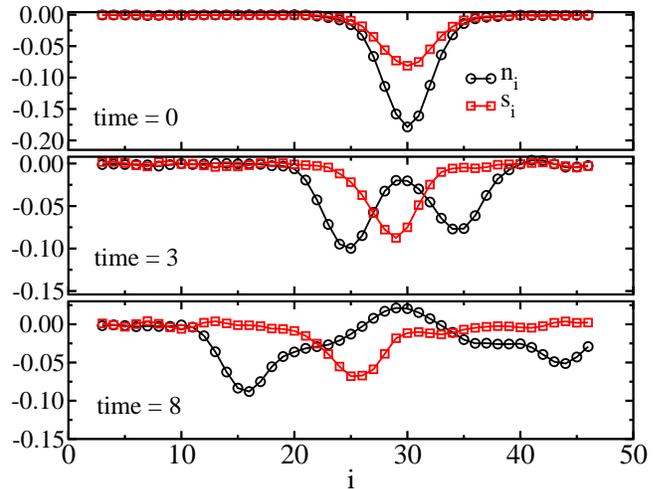}
\caption{Charge (circles) and spin (squares) densities in the strong-coupling limit ($U = 8t_h$ and $V=0$). The wave packet, with charge and spin, is created at $t=0$ centered at $j_0=30$. SCS and charge fractionalization are observed before $t=3$. The spin velocity is much smaller than that of the charge. Note that the spin excitation is almost completely localized (see Eqs.~(\ref{vsSC}) and (\ref{vcSC}) with $V=0$). Charge fractionalization is stronger than in the weak-coupling case and it approaches the behavior expected for the strong-coupling regime (see text).}
\label{Fig3}
\end{figure}

Figure~\ref{Fig1} shows snapshots of $n_i$ and $s_i$ for $U/t_h=2$ and $V=0$ at different times. At $t=0$, a wave packet with charge and spin, is created centered at $j_0=30$. The spin and charge move in the same direction but at different velocities (see Eqs.~(\ref{vsWC}) and (\ref{vcWC}) with $V=0$). SCS and charge fractionalization, as predicted by the LL theory, are evident starting at $t=3$. The charge splits into left- and right-moving packets. Note, however, that the fractionalization is small and most of the charge moves to the left, as expected for small $U$. LL theory predicts  that charge fractionalizes according to the formula $q = q_L + q_R$ where $q_L=q(1+K_\rho)/2$ and $q_R=q(1-K_\rho)/2$ are the left- and right-moving charges, respectively, and $K_\rho=(1+U/\pi v_F)^{-1/2}$ is the LL parameter. For small $U$, $K_\rho \approx 0.87$ and most of the charge moves in one direction. Using Eqs.~(\ref{vsWC}) and (\ref{vcWC}), the ratio of the velocities in this limit is $v_c/v_s \approx 1+U/\pi v_F \agt 1$ in agreement with the
numerical results in Fig.~\ref{Fig1}.

Figure~\ref{Fig2} shows the results for $U/t_h = 2$ and different values of $V$. For each $V$, $n_i$ and $s_i$ are shown at times $t=0$ and $t=6$. For $V/t_h \le 0.6$, both $v_c$ and $v_s$ increase with $V$ as predicted by the weak-coupling analysis based on field-theoretical formulas of Sec.~\ref{Weak Coupling}. Note, however, that $v_c$ has a stronger dependence on $V$ than $v_s$ and SCS is enhanced as shown by Eqs.~(\ref{vsWC}) and (\ref{vcWC}). Indeed, the ratio $v_c/v_s \approx 1+(U+2V)/\pi v_F$ shows explicitly an increase of the SCS in the weak-coupling regime. It is interesting to point out that, in spite of the effect on SCS, the charge 
fractionalization is weakly affected by $V$. This trend can be explained by looking at the bosonization equations for $q_L$ and $q_R$. In this regime and for the half-filled case, $K_\rho = \left[1+(U+6V)/\pi v_F\right]^{-1/2}$. For $U = 2t_h$, the numerical value, obtained from $K_\rho$, of the left-moving charge fraction is $q_L/q \approx 0.94$ for $V=0$. This fraction decreases slightly to $q_L/q \approx 0.85$ for $V=0.8t_h$. 


\begin{figure}
\includegraphics*[width=\columnwidth]{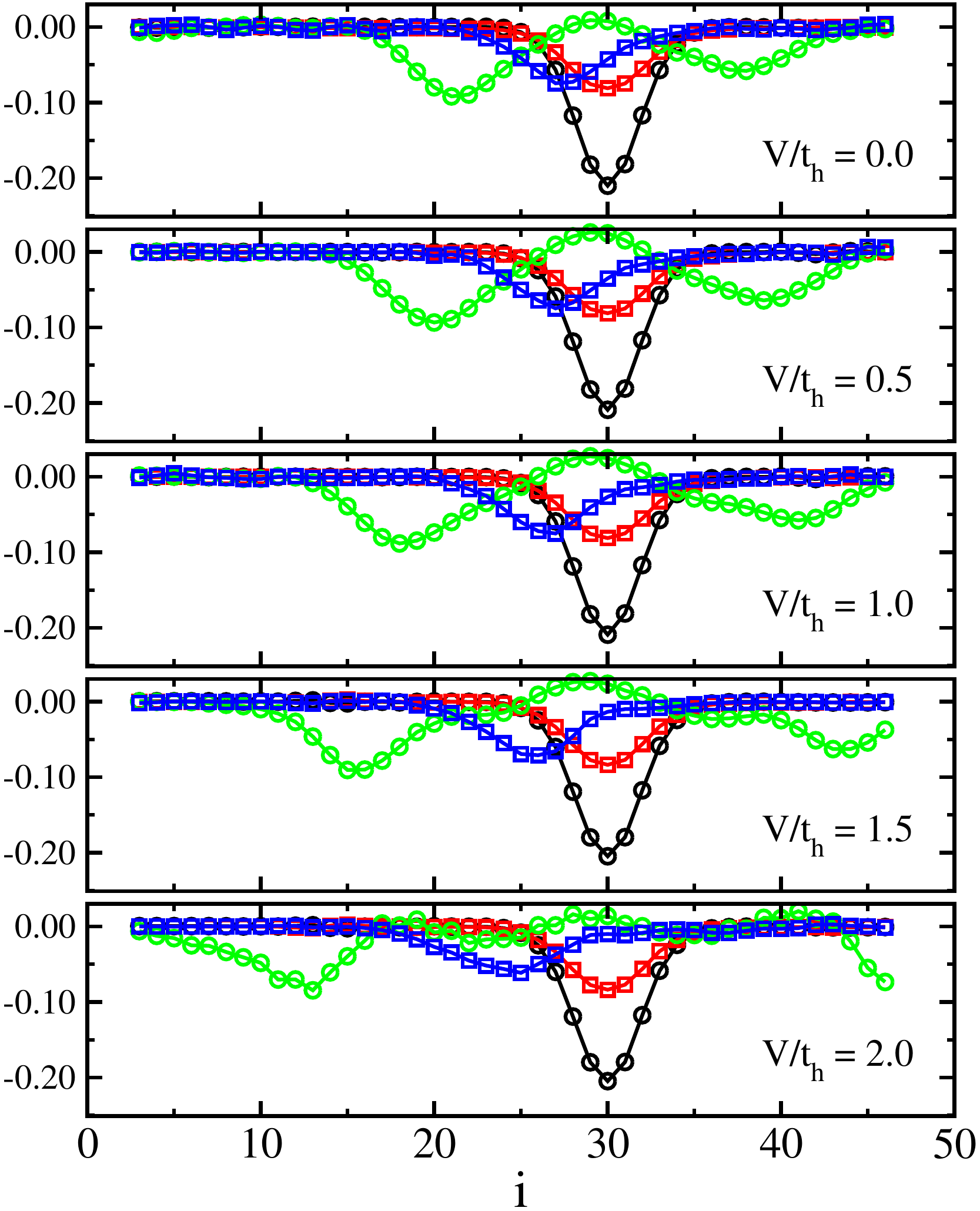}
\caption{Charge and spin densities, $n_i$ (circles) and $s_i$ (squares) for large $U$, $U=8t_h$, and different values of $V$. For each $V$, the results are shown at $t=0$ and $t=5$. The spin velocity is small and increases slowly with $V$ as shown by Eq.~(\ref{vsSC}). On the other hand, the charge velocity exhibits a strong dependence on $V$, see Eq.~(\ref{vcSC}). Note that increasing $V$ from 0 to $2t_h$ leads to a substantial enhancement of the SCS. The charge fractionalization, however, is weakly affected.}
\label{Fig4}
\end{figure}

The results for large $U$, more specifically $U=8t_h$ and $V=0$, are shown in Fig.~\ref{Fig3}. SCS and charge fractionalization are observed in the strong-coupling limit as well. This is shown in snapshots starting at $t=3$. The spin velocity is much smaller than that of the charge, as shown by Eqs.~(\ref{vsSC}) and (\ref{vcSC}). SCS is clearly seen at even earlier times. In the strong-coupling limit this may be related to the velocity ratio $v_c/v_s \approx t_h/J \gg 1$, which is remarkably larger than in the weak-coupling case. Charge fractionalization is also more pronounced in this limit. The left- and right-moving charge fractions are comparable. In other words, the charge splits into almost equal fractions moving in opposite directions. In the limit $U/\pi v_F\gg 1$ and close to half-filling, as in this case, the relevant low-energy physics is dictated by the motion and effective interaction of holes. A crossover in the correlations of these holes between power-law and long-range behavior is expected depending on the energy scale. This crossover is associated with the opening of a gap in the charge sector at half-filling. If the correlations are long-ranged, one expects an effective correlation exponent close to zero giving rise to an almost even charge fractionalization (see Eq.~(3.10) in Ref.~\onlinecite{Schulz94}).
 
Figure~\ref{Fig4} shows the charge and spin densities, $n_i$ and $s_i$ for the strong-coupling limit ($U=8t_h$) and different values of $V$. For each $V$, the results are shown at $t=0$ and $t=5$. The spin velocity is small and increases slowly with $V$ in agreement with Eq.~(\ref{vsSC}). The charge velocity also increases with $V$ displaying a stronger dependence than that of $v_s$. This is in agreement with the strong-coupling analysis presented in Sec.~\ref{Strong Coupling} (see Eq.~(\ref{vcSC})). As a consequence of this strong dependence on the NN repulsion, SCS is therefore enhanced. The velocity ratio of the excitations is $v_c/v_s \approx (v_F/\pi J)(1+2V/\pi v_F)$ leading to a substantial increase of the SCS. The charge fractionalization, however, is very weakly affected as can be expected from the value of the correlation exponent for the spinless fermions model, $K_\rho \approx [1+ \frac{2}{\pi}\frac{V-t^2_h/U}{v_F}(2-\frac{\pi^2\delta^2}{2})]^{-1/2}$. This equation is derived using the same reasoning as in Sec. \ref{Strong Coupling}.

For the large $V$ limit, the CDW fluctuations become dominant. In the half-filling regime and for $V>U/2$, the ground state is a band insulator with alternating empty and doubly-occupied sites. As expected for band insulators, SCS is suppressed. An electron (or hole) added in the system moves with an effective hopping $\tilde t \propto t_h^2/(V-U/2)$. In this regime, $v_c \approx v_s$ and both velocities are proportional to $\tilde t$. This has been observed in the numerics (not shown) for large $V$.

It is also worthwhile to point out that the presented results can be extended to models with a similar GS phase diagram such as the Hubbard model with an ionic potential $\Delta$.~\cite{Tincani} This model at half-filling also exhibits an SDW for $U>\Delta$ where SCS and charge fractionalization are expected. For $U<\Delta$, a CDW develops and therefore there is no SCS or fractionalization. In the vicinity of $U\sim\Delta$, the system shows a spin dimerized phase similar to the bond-ordered phase of the extended Hubbard model.

\section{Conclusions}
\label{Conclusions}

Using tDMRG, we have studied the spin-charge separation in the extended Hubbard model focusing on the effect of nearest-neighbor repulsion. The excitation, introduced in the system by creating a Gaussian wave-packet, carries charge and spin. 
We have studied both the weak- and strong-coupling regimes. In both regimes, the effect of $V$ is to enhance spin-charge separation 
tendencies. For small and intermediate $V$, both spin and charge velocities increase with $V$, but charge velocity has a stronger dependence, and spin-charge separation is more pronounced. As $V$ is increased further, and the ground state approaches the CDW regime, the ratio of the velocities $v_c/v_s\sim 1$ and spin-charge separation and charge fractionalization are suppressed.

Notwithstanding the qualitative similarity between results in the weak- and strong-coupling regimes, their explanations differ. The results in the weak-coupling regime can be explained using bosonization equations. In the strong-coupling regime, on the other hand, the results are better understood by mapping the Hamiltonian into an effective low-energy model, and studying the spin and charge sectors independently. Our numerical results also show that spin-charge separation goes beyond the asymptotically low-energy theory of LLs for the extended Hubbard model. We would like to remark the weak effect of $V$ on charge fractionalization. Both in the weak- and strong-coupling regimes, charge fractionalization increases only slightly as $V$ increases. Similar results (not shown here) are obtained in the intermediate coupling regime, $U \approx 4t_h$.

tDMRG should enable future work extending the present study to include more complicated geometries and systems, such as ladders, multiorbital models,\cite{Onishi} and materials with 
strong electron-phonon coupling.

\begin{acknowledgments}
The authors thank C. D. Batista for insightful discussions. This work was supported by the Center for Nanophase Materials Sciences, sponsored by the Scientific User Facilities Division, Basic Energy Sciences, U.S. Department of Energy, under contract with UT-Battelle. K.A., J.R. and G.A. acknowledge support from the DOE early career research program. E.D. is supported in part by the U.S. Department of Energy, Office of Basic Energy Sciences, Materials Science and Engineering Division.
\end{acknowledgments}

\end{document}